\documentstyle[12pt, epsfig]{ioplppt}
\begin{document}

\title{Glassy dynamics in granular compaction}

\author{Anita Mehta}
\address{S N Bose National Centre for Basic Sciences\\
Block JD Sector III Salt Lake\\
Calcutta 700091, India\\
email: anita@boson.bose.res.in}
\author{
G C Barker}
\address{
Institute of Food Research,\\
Norwich Research Park,\\
Colney Lane, Norwich NR4 7UA, UK\\
email: barker@bbsrc.ac.uk}
\begin{abstract}
Two models are presented to study the influence of slow dynamics on
granular compaction. It is found in both cases that high values
of packing fraction are achieved only by the slow relaxation of
 cooperative structures. Ongoing work to discuss the full implications
of these issues is discussed.

\end{abstract}


\submitted{{\noindent \it 30 November 1999  }}

\section{Introduction}

The importance of glassy dynamics in granular media was recognised
well before recent experiments
\cite{knight}, \cite{nowak},
\cite{nowak1}
 in granular compaction
\cite{rmp}
 made some of the
underlying ideas concrete \cite{dg}, \cite{sam}. In particular, the idea that
{\it two} dynamical mechanisms 
are needed to explain some of the observed experimental phenomena \cite{sidprl}
has previously been put forward \cite{amphys}; a cooperative mechanism
embodies the slow dynamics of relaxing granular clusters, while
a single-particle mechanism represents mobile grains
moving between clusters.

In this paper, we discuss the results from two models that  
incorporate glassy dynamics into the phenomenon
of granular compaction. Our first approach is based on
a hybrid Monte Carlo dynamics
(which contains both single-particle and cooperative 
components), and the second uses a cellular automaton model,
that includes, in addition to the canonical threshold-driven grain
flow, a representation of cooperative reorganisation
via dynamical disorder. In both cases, we find that a driven
sandpile undergoes compaction largely as a result of the slow
dynamics of cooperative motion.

While it has been shown in earlier work that granular compaction
beyond values of packing fraction of 0.56 occurs almost
entirely because of the cooperative relaxation of grain clusters \cite{PRL},
the present results from Monte Carlo simulations
shed some light on the details of this
compaction in relation to experiments
 \cite{knight}, \cite{nowak}.
Equally, on realising that the analogue  of bulk compaction
corresponds to smoothing of the sandpile surface, we demonstrate
that this is indeed what occurs in our driven and dynamically
disordered cellular automaton model.

\section{Bulk compaction: a hybrid Monte Carlo model}
\label{sec:section1}

Recent experiments 
 \cite{knight}, \cite{nowak}, \cite{nowak1}
have demonstrated the importance of glassy dynamics
in granular compaction. In ref. \cite{knight},
the authors 
 observed a monotonic increase 
of packing fraction with excitation intensity.
In refs. \cite{nowak}, \cite{nowak1}, a more complex (and  now
well-established) behaviour was observed:
an initial ramping up of the intensity led, as before, to
an increase of volume fraction. However at a certain
point (the 'irreversibility point' )
any subsequent increase of volume fraction could only
be generated by {\it decreasing} the intensity of vibration.
The first of these two regimes, called
the irreversible branch, was interpreted
as the increase in packing fraction resulting
from the shaking out of
 voids 
from an initial, loosely packed state. The second regime,
called the reversible branch,
is in accord with the earlier predictions
of Monte Carlo simulations \cite{PRL}, \cite{PRA}; one of the features of
 the reversible branch, as its name implies, is that
 evolution in the directions of either increasing
or decreasing excitation intensity yields reproducible results in packing
fraction.  Monte Carlo simulations  
are predicated on
 reversible transitions between configurations,
so that their predictions lie entirely on the reversible
branch of the experimental curves.

In this paper, 
we take these investigations further in a bid
to understand the theoretical implications of
the experimental phase diagram.
 We find a transition point in the behaviour of a
shaken sphere packing; for a range of shaking
intensities, there is a transition to an ordered close-packed 
state (with packing fractions $\phi \ge  0.61$) after sufficiently
long shaking times. For intensities below this range the
powder remains stuck in some configurations and therefore cannot
crystallise: for intensities above this range, the behaviour
is analogous to 'quenching' and crystallisation is therefore
inhibited. We argue that the lower bound for this
range of shaking intensities corresponds to 
the 'irreversibility point' observed in experiments \cite{nowak}.

Our simulations use uniform hard spheres, subjected to
non-sequential
 reorganizations which represent the effect of shaking.
A  variable  shaking  amplitude $A$ is parametrised
in units of the particle diameter; thus for example,
$A=1.0$ means that shaken particles
are able to move longitudinally and laterally
by, on average, one particle diameter (subject to volume exclusions) during
a shake cycle. 
  The details
of the shaking algorithm have been  discussed elsewhere
(\cite{PRL},\cite{ambook}).
Briefly,  one cycle of vibration of the granular
assembly (corresponding to one timestep of our simulation) is modelled by :
\begin{enumerate}
\item a vertical dilation
of the granular bed, in proportion
to the shaking amplitude $A$ 
\item a stochastic
rearrangement 
 of the individual particles in transverse
directions, with available free volume proportional
once again to the shaking amplitude
\item and finally a {\it cooperative} recompression of the assembly
as each grain lands on the substrate alone or with neighbours; in the latter
case, arches would form.
\end{enumerate}

In conventional Monte Carlo, the cooperative step is absent,
and particle reorganisation is sequential; this corresponds
to a regime of 'fast' dynamics, driven by the inertia of
the grains. In contrast our simulations interpolate naturally from this
regime to one with slow dynamics, characteristic of that found
in glassy motion, because of the inclusion of  cooperative
rearrangements in the last step. In regions
where the shaking amplitude $A$ is large, one can
discuss the dynamics in terms of the motion
of the individual particles, since any arches that
form at one time step are rapidly destroyed at another, i.e.
there is little indication of long-lived cooperative motion.
In contrast, for low shaking amplitudes, the cooperative
step which we have added to our Monte Carlo procedure is crucially
important for modelling the correlated motion of grains that is important
in slow dynamics. In earlier work we have studied this slow dynamics using
 displacement correlation
functions, and thus defined the concept of a dynamical cluster \cite{PRA}. 
Similar displacement correlations have  subsequently been
studied in the context of glasses \cite{sharon}.
This crossover between the fast and slow dynamics can be viewed also in terms of the interpolation between two
 effective temperatures
in a granular medium, the first corresponding to the conventional
granular temperature defined in terms of the inertia
 of the grains \cite{savage},
and a second corresponding to a density-related temperature first
defined by Edwards \cite{sam1}, and known as the compactivity. The details
of such interpolation are discussed analytically 
elsewhere \cite{ars};  a  recent  approach
which looks at the issue of two temperatures at a more microscopic level
is due to Kurchan \cite{jorge}.

We now present our simulation results.
Starting from a random loose packing with packing fraction $\phi \sim 0.54$,
 a fixed shaking intensity
leads to packings with  steady state values for the packing fraction. The
steady state is approached after short
or long transients,  depending on the value of shaking intensity. 
However, within a range of excitation intensities, further shaking for extended
periods may produce
 a jump
to an ordered close-packed state; we term this shaking-induced crystallisation. Outside
this range, we have not observed crystallisation, at least for our simulation
times, though we speculate that for extremely long times, and for shaking
intensities {\it below} our range, a jump to the crystalline
state could be a possibility (see below).

Our simulations are carried out in an 8x8x8 periodic box
filled with monosize unit spheres; there are approximately 700 spheres
in total. Figure 1  shows the variation of the packing fraction with
time t (in shaking cycles), as the spheres are shaken, 
 at amplitudes $A=0.05, 0.5$ and $1.2$ respectively.
For $A = 0.5$, (Figure 1b), we notice a sharp rise in packing
fraction to about $\phi = 0.68$  at $t \sim 900$. This does not
happen in the other two cases, at least for our times of observation. 
(At the lowest shaking intensity we have followed the time
series for $2.10^5$ cycles).
For large shaking amplitudes (Figure 1c) the dynamics
is akin to that of fluidisation, while for very small
$A$, the granular bed appears to be stuck in 'supercooled'
configurations.
These results  indicate  that there is a range of amplitudes
 that  will allow
for the granular assembly to crystallise; clearly
this depends on the observation time, since, at very low shaking intensity,
 we cannot
rule out a jump to the 
near-crystalline state from one of the
supercooled states 
over infinitely long times. 

\begin{figure} 
\hbox to\hsize{\epsfxsize=0.3\hsize
{\epsfbox{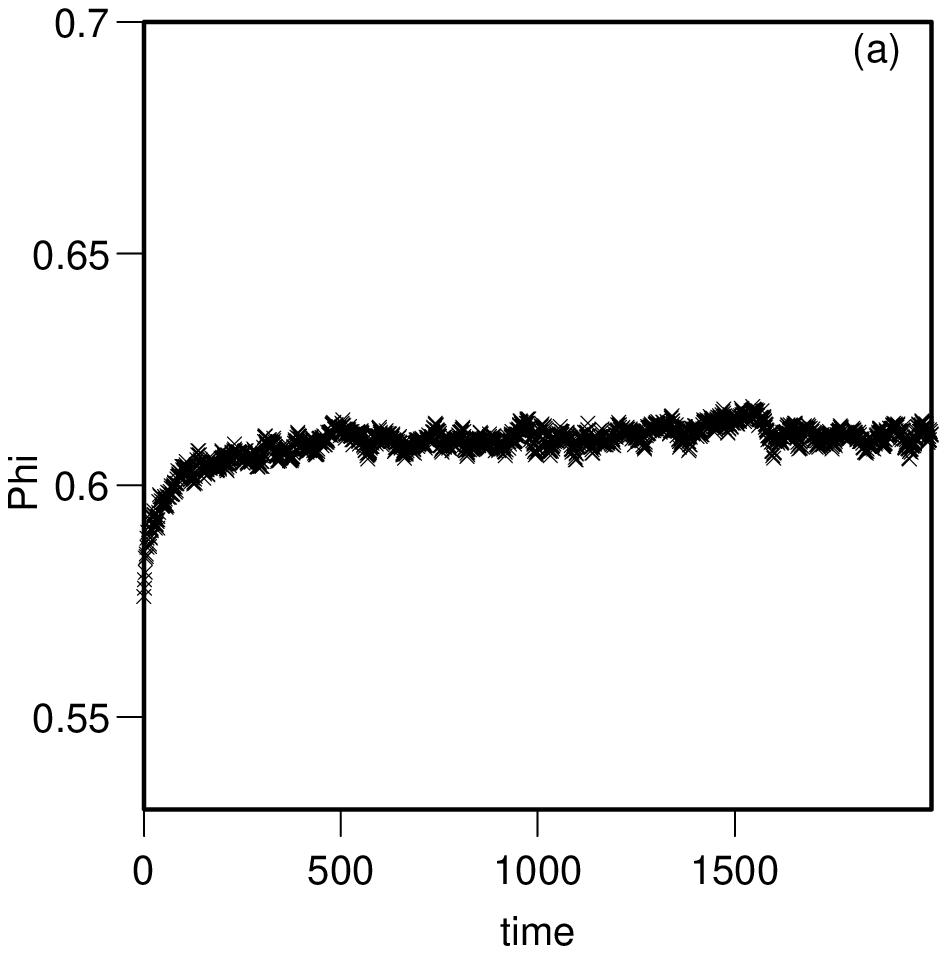}}
{\epsfbox{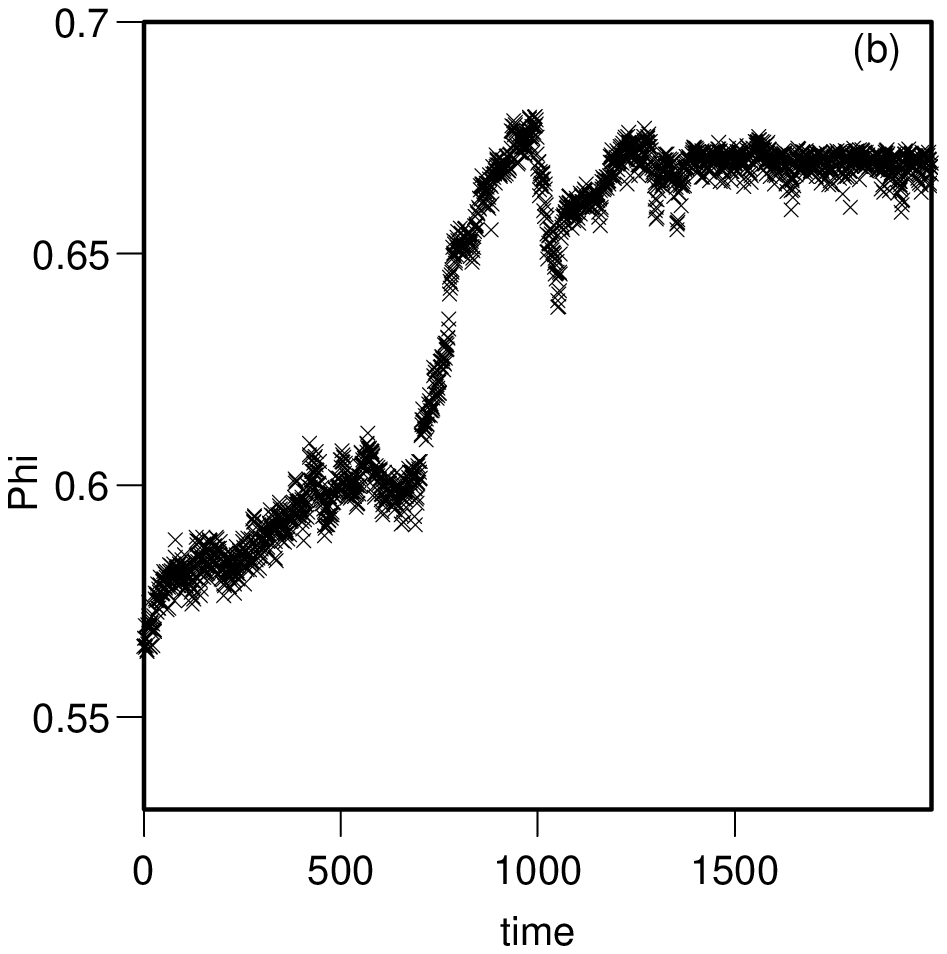}}
{\epsfbox{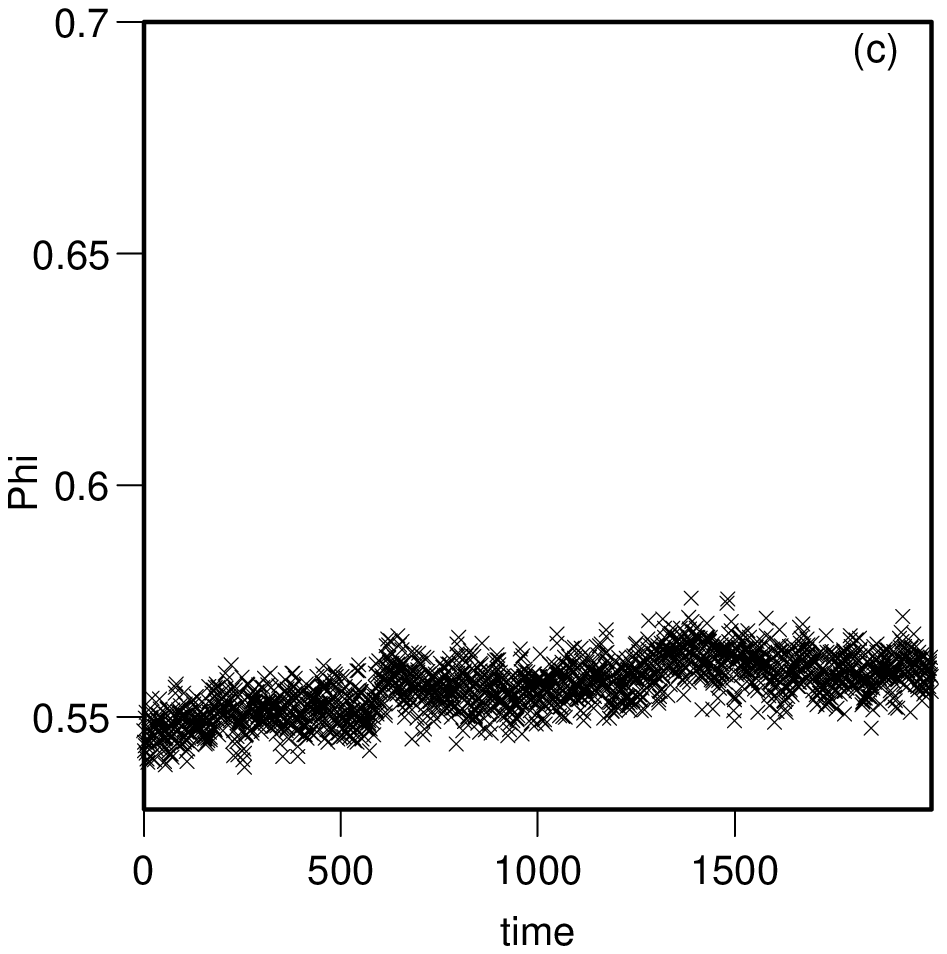}}}
\caption{Plots of packing fraction $\phi$ vs time t
for a) $A=0.05$ b)$A=0.5$ c)$A=1.2$. Note the approach to crystallisation
in fig. 1b}
\label{fig1} 
\end{figure} 

In Figure 2, we show clusters of approximately
 300 spheres 
generated  at $t = 2000$, corresponding respectively to
$ A=0.05 $
and $ A=0.5$. In the latter case, the snapshot  is taken
after the   ordering transition. The
structures are  fundamentally different, leading
to our conjecture that the latter corresponds to an ordered state.

\begin{figure} 
\hbox to\hsize{\epsfxsize=0.4\hsize
{\epsfbox{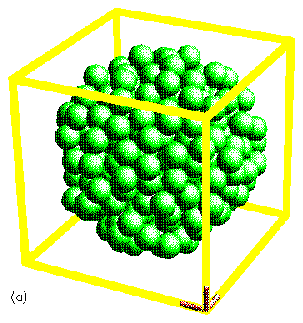}}
{\epsfbox{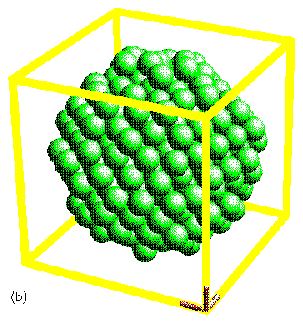}}}
\caption{An example of typical clusters obtained after 2000 timesteps
for a) $A=0.05$ b)$A=0.5$. Note the crystalline-like ordering in the second case}
\label{fig2} 
\end{figure} 

These results lend weight to a conjecture that, as in glasses,
a barrier height distribution \cite{ars}  exists between configurations;
 at very low intensities of vibration, the driving force 
cannot typically force the system to cross barriers,
 leading to 'stuck' configurations in the powder,
which look supercooled.These barriers are entropic,
and the barrier height represents the energy threshold
needed to move the system from one configuration to another.
The relaxation rates both for fast and slow dynamics, are expected 
to be activated processes over the (random) distribution of
barrier heights; in each case, the appropriate effective temperature 
controls the kinetics \cite{ars}.

 As the intensity
increases, the configurations can evolve more easily
(the powder becomes more ergodic)
so that in principle, for infinite times,
the system can achieve a crystalline limit.
In order to understand why little compaction
occurs (and why certainly the transition to the crystalline
state would be most unlikely) above the allowed range, we 
must consider the mechanism of compaction discussed 
previously \cite{PRA}. As the intensity of vibration
decreases, fewer and fewer grains are able to break away
from their clusters, so that structures such
as arches are long-standing even during driving. 
The relaxation of these arches in a cooperative sense (measured
by displacement correlation functions \cite{PRA}) leads
to the decrease of the  void space  which is trapped in the arches ("bridge collapse")
and overall, the powder compacts. In contrast,
arches make and break during strong driving (the autocorrelation
function for grains decays to zero rather rapidly \cite{PRA})
so there are strong fluctuations in the total volume of void space 
 and no overall compaction results. (Another way
of seeing this is that the effective temperature
for such fast dynamics affects the inertia of individual
grains, rather than helping the system overcome
configurational barriers en route to compaction). This fluctuating
behaviour in  the global packing fraction of the powder
can be seen very clearly from Fig. 1c, and was also noted in
ref. \cite{nowak1}.

Further work is in progress to refine and extend these
results, but we discuss below the possible relationship between the simulations and 
 the experiment in refs. \cite{nowak}, \cite{nowak1}.
There are
 important differences between our simulations
and the compaction experiments.
The irreversible branch of the Chicago experiments consists
of transitions between states that
can only be reached in one direction - 
this directionality corresponds to the ramping up of the
reduced acceleration. The trajectories of our simulation
on the other hand,  connect what we believe to
be "equilibrium states" (at least within our observation times)
and transitions are in general  reversible.
This observation \cite{nowak} indicates that our simulation data
correspond to the reversible branch of the experimental graphs.
Another important difference concerns the fact that our
data  represent the time evolution of the packing
fraction of the powder at a fixed shaking intensity, while
that of the Chicago experiments concern the evolution of the packing
fraction as a function of the shaking intensity, and of the {\it time}
spent at each intensity. It was noted \cite{nowak1} that, except
at very low intensity driving, with the consequent preponderance
of 'supercooled' states in the powder, it may be possible to 
reach 
the reversible branch by an extended time of excitation, at a fixed intensity.

 Our results include this interesting possibility. While clearly
the "crystallization" observed in Fig 1b might be regarded as an
irreversible transition
 (it is easier for a close-packed array to be shaken
down to lower density than for a loose-packed powder to make a sudden
jump -  overcoming a configurational barrier  - to a crystalline density),
 it is included in a trajectory that has
reversible
steps.
On the other hand, for extremely low intensities, it is likely that
the super-cooled configurations of the powder would be persistent on
the time scale of experiments. We speculate therefore that the lower bound
for the range of intensities where such crystallisation is possible
corresponds to the irreversibility point observed in experiments \cite{nowak},
\cite{nowak1}.

In ongoing work, we are examining 
the influences of increased observation times
at fixed, low intensity shaking, on the point at which
the transition to crystallinity is observed. In addition we are making an 
accurate determination of the range of intensities, for
different system sizes and times of observation, for which
this crystallisation is possible and we are mimicking
the experiment by looking at varying shaking intensities at fixed ramp
rates.  We hope in this way to determine the value of
 the  irreversibility point, as well as the behaviour of the system around it.

\section{Surface compaction: a disordered cellular-automaton model}
\label{sec:section2}

We now examine the issue of 'surface compaction', or
smoothing of a driven sandpile surface.  
As deposition occurs on
a sandpile surface, clusters of grains grow unevenly at different
positions and roughness builds up until further deposition renders
some of the clusters unstable. These then start 'toppling', so
that grains from an already unstable cluster flow down the
sandpile, knocking off grains from other similar clusters which
they encounter. The net effect of this is to 'wipe off'
protrusions (where there is a surfeit of grains at a cluster) and
to 'fill in' dips, where the oncoming avalanche can disburse some
of its grains. In short, the surface is smoothed by the passage of
the avalanche so that there is a rough precursor surface, and a
smoothed post-avalanche surface.

 We
have used a cellular-automaton model (CA) \cite{amca} of an evolving sandpile
to examine this issue; this model appears  \cite{curr}
to be the discrete version of an earlier continuum model \cite{amcoupled1}.
This CA model is a
'disordered'  version of the basic Kadanoff cellular
automaton \cite{kadanoff}; a further degree of freedom, that
involves granular reorganisation within columns, is added to the
basic model which includes only granular flow between columns.
As in the previous section, this extra ingredient of intra-column
reorganisation is a way to introduce slow cooperative dynamics into
the system. As we will see, these orientational relaxations cause surface
smoothing of our CA sandpile, mirroring the way
 in which the cooperative step in the 
Monte Carlo caused the observed bulk compaction (see above).

Our disordered model sandpile \cite{caparam} is built from
rectangular lattice grains, that have aspect ratio $a \le 1$,
arranged in columns $i$ with $ 1 \le i \le L$, where $L$ is the
system size. Each grain is labelled by its column index $i$  and
by an orientational index $0$ or $1$, corresponding respectively
to whether the grain rests on its larger or smaller edge. The two
grain orientations represent regions of either loose (type $1$)
or close (type $0$) packed material.

The dynamics of our model have been described at length elsewhere
\cite{amca}; briefly,
\begin{itemize}
\item
Grains are deposited on the sandpile with fixed probabilities 
for orientation in the $0$ or $1$ states.
\item The incoming grains,
as well as all the grains in the same column, can then 'flip' to
the other orientation stochastically (with probabilities which
decrease with depth from the surface). This 'flip', or change of
orientation, is our simple representation of {\it collective
dynamics in granular clusters} since typically clusters reorganise
owing to the slight orientational movements of the grains within
them \cite{ambook}. The transition probabilities in this case
involve
scale heights which are weighted so as to favour the destruction
of voids, as in a slowly consolidating granular material \cite{ball}.
\item
The height of column
$i$ at time $t$, $h(i,t)$, can be expressed in terms of the
instantaneous numbers of $0$ and $1$ grains, $ n_0(i,t)$ and
$n_1(i,t)$ respectively:
\begin{equation}
        h(i,t) = n_1(i,t) + a n_0(i,t)
\end{equation}
\item
Finally, grains fall to the next column down the sandpile
(maintaining their orientation as they do so) if the height
difference exceeds a specified threshold in the normal way
\cite{kadanoff} (the pile is local, limited and has a fall number
of two). At this point, avalanching may occur.
\end{itemize}

We begin with the 
principle of  dynamical scaling for sandpile cellular
automata \cite{curr} in terms of the surface width $W$ of the sandpile
automaton:

\begin{eqnarray}
 W(t) &\sim&  t^{\beta},  t \ll t_{crossover} \equiv L^z
\label{betaeqn}\\
 W(L) &\sim&  L^{\alpha}, L \to \infty \label{alphaeqn}
\end{eqnarray}

As in the case of interfacial widths, these equations signify the
following sequence of roughening regimes:
\begin{enumerate}
\item To start with, roughening occurs at the CA sandpile surface
in a time-dependent way; after an initial
transient, the width scales asymptotically with time $t$ as $t^{\beta}$,
where
$\beta$ is the {\it temporal roughening} exponent. This regime is
appropriate for all
times less than the crossover time $t_{crossover} \equiv L^z$,
where $z$
= $\alpha/\beta$
is the dynamical exponent
and $L$ the system
size.
\item After the surface has {\it saturated}, i.e. its width no longer
grows with time, the {\it spatial roughening} characteristics of
the mature interface can be measured in terms of $\alpha$, an
exponent characterising the dependence of the width on  $L$.
\end{enumerate}

We define the  surface width $W(t)$ for a sandpile automaton in
terms of the mean-squared deviations from a suitably defined mean
surface; in  analogy with the  conventional counterpart for
interface growth \cite{krug}, we define the instantaneous mean
surface of a sandpile automaton as the surface about which the sum
of column {\it height} fluctuations vanishes. Clearly, in an
evolving surface, this must be a function of time; hence all
quantities in the following analysis  will be presumed to be
instantaneous.

The mean slope $<s(t)>$ defines  expected column heights,
$h_{av}(i,t)$, according to
\begin{equation}
            h_{av}(i,t) = i <s(t)>
\end{equation}
where we have assumed that column $1$ is at the bottom of the
pile. Column height deviations are defined by
\begin{equation}
        dh(i,t) = h(i,t) - h_{av}(i,t)  =  h(i,t) - i <s(t)>
\end{equation}
The mean slope must therefore satisfy
\begin{equation}
            \Sigma_i[ h(i,t) - i <s(t)> ] = 0
\end{equation}
since the instantaneous deviations about it vanish; thus
\begin{equation}
            <s(t)> = 2 \Sigma_i[h(i,t)] / L(L+1)
\end{equation}

The instantaneous width of the surface of a sandpile automaton,
$W(t)$, can  be defined as:
\begin{equation}
            W(t) = \sqrt{ \Sigma_i[dh(i,t)^2] / L }
\end{equation}
which can in turn be averaged over several realizations to give,
$<W>$, the average surface width in the steady state.

\begin{figure}
\hbox to\hsize{\epsfxsize=0.25\hsize
{\epsfbox{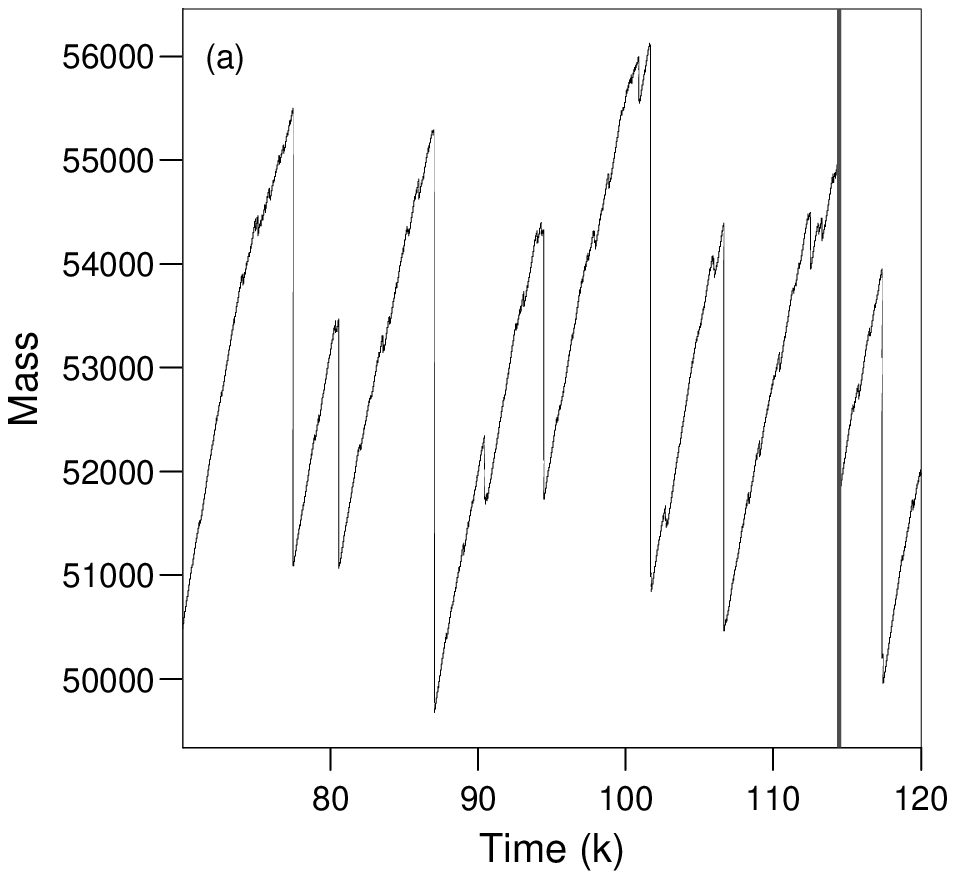}}
{\epsfbox{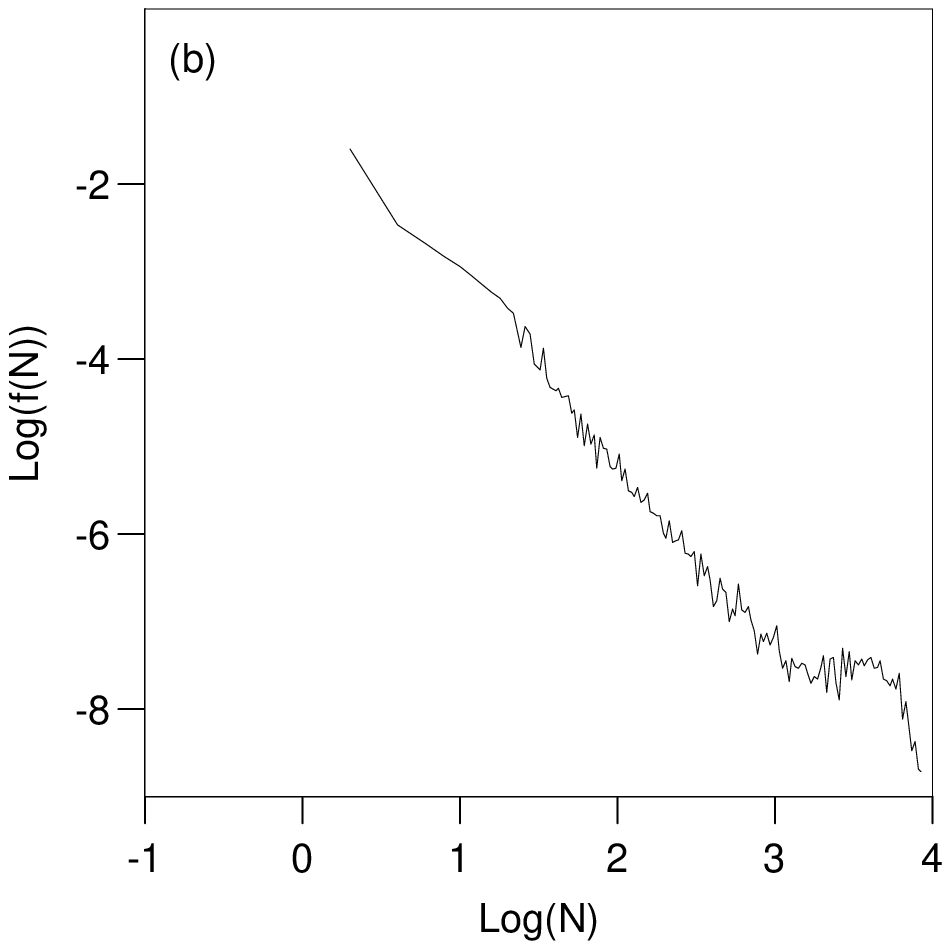}}
{\epsfbox{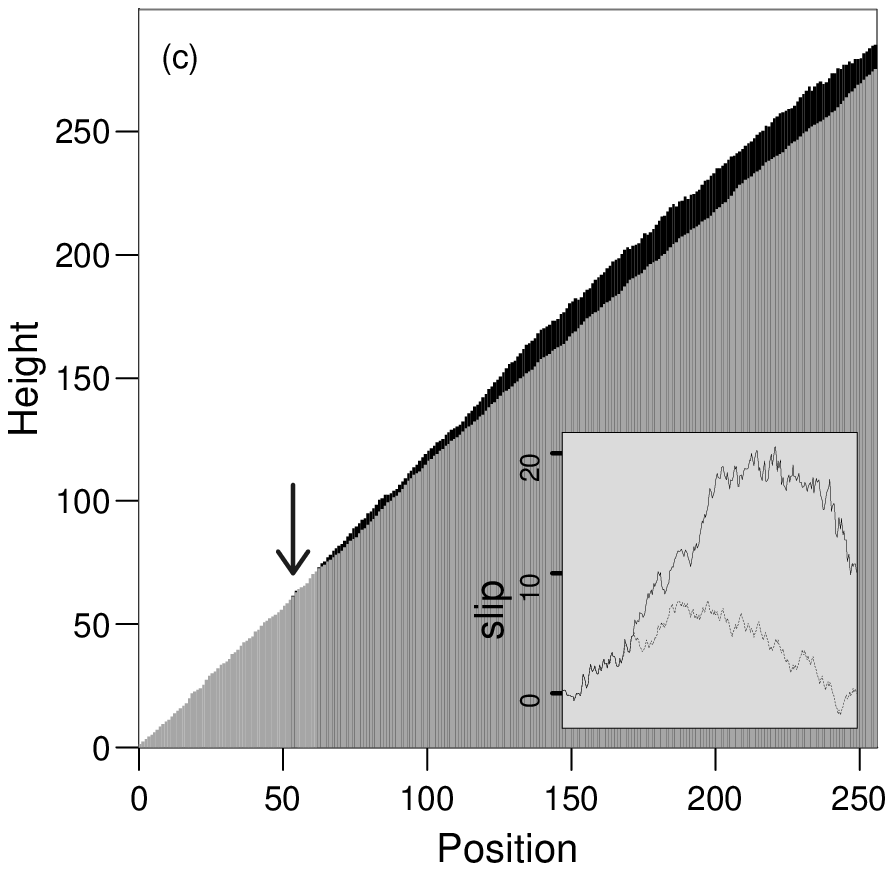}}
{\epsfbox{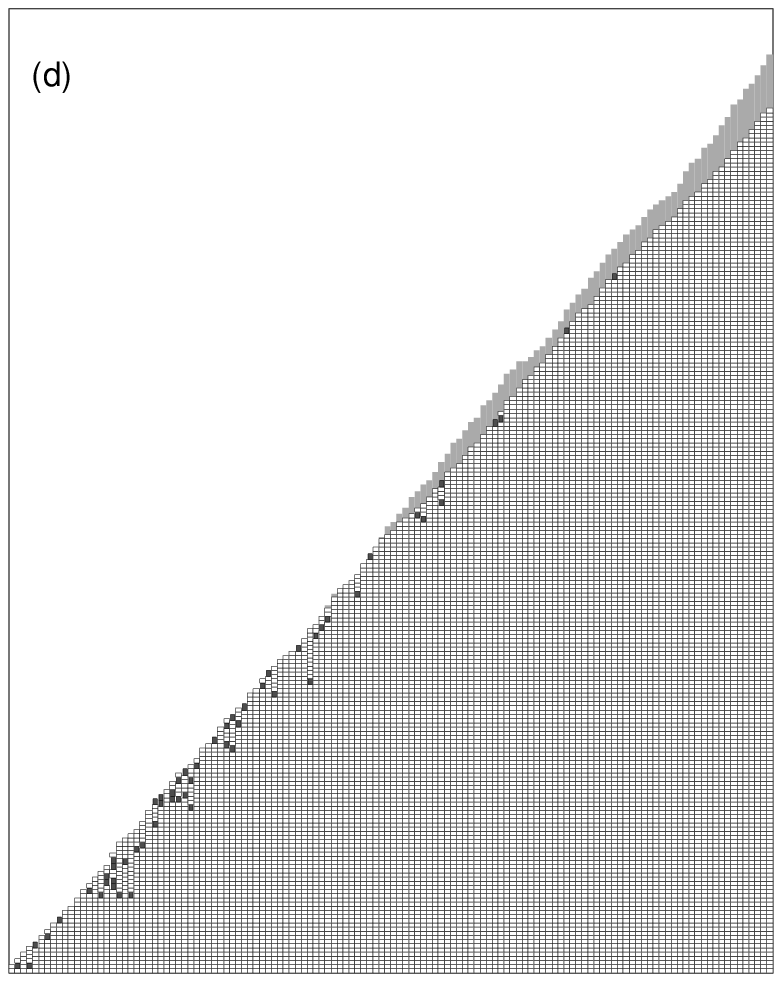}}}
\caption{(a) A time series of the mass for a model sandpile
($L=256$) that has been built to include a surface layer
containing structural disorder. The vertical line indicates the
position in this series of the large avalanche illustrated in
Figure 3 c, d. (b) A log-log plot of the event size
distribution for a model sandpile ($L=256$) that includes a
surface layer containing structural disorder. (c) An illustration
of a large wedge shaped avalanche in a model sandpile ($L=256$)
that has been built to include a surface layer containing
structural disorder. A lighter aftermath pile has been superposed
onto the dark precursor pile and an arrow shows the point at which
the event was initiated. The inset shows the relative positions of
the two surfaces and their relationship to a pile that has a
smooth slope. (d) A detailed picture of the internal structure of
a model sandpile in the aftermath of a large avalanche event. The
individual grains of the aftermath pile (for columns $1-128$ of a
sandpile with $L=256$) are superposed on the gray outline of the
precursor pile.} 
\label{fig3} 
\end{figure}
Figure 3(a)  shows a time series for the mass of a large
($L=256$) evolving disordered sandpile automaton. The series has a
typical quasiperiodicity \cite{held}. The vertical line denotes
the position of a particular 'large' event, while Figure
3(b) shows the avalanche size distribution for the
sandpile. Note the peak, corresponding to the preferred large
avalanches, which was analysed extensively in earlier work
\cite{amca}. Our data shows that the avalanche highlighted in
Figure 3(a) drained off approximately $5$ per cent of the
mass of the sandpile,  placing it close to the 'second peak' of
Figure 3(b). Figure 3(c) shows the outline of
the full avalanche before and after this event with its initiation
site marked by an arrow; we note that, as is often the case in one
dimension, the avalanche is 'uphill'. The inset shows the relative
motion of the surface during this event; we note that the
signatures of smoothing by avalanches are already evident as the
precursor state in the inset is much rougher than the final state.
Finally we show in Figure 3(d) the grain-by-grain picture
of the aftermath pile superposed on the precursor pile, which is
shown in shadow. An examination of the aftermath pile and the
precursor pile shows that the propagation of the avalanche across
the  upper half of the pile has left only a very few disordered
sites in its wake (i.e. the majority of the remaining sites are
$0$ type) whereas the lower half (which was undisturbed by the
avalanche) still contains many disordered (i.e. $1$ type) sites in
the boundary layer. This leads us to suggest that the larger
avalanches rid the boundary layer of its disorder-induced
roughness, a fact that is borne out by our more quantitative
investigations.
In fact, our studies have revealed that the very largest
avalanches, which are system-spanning, remove virtually all
disordered sites from the surface layer; one is then left with a
normal 'ordered' sandpile, where the avalanches have their usual
scaling form for as long as it takes for a layer of disorder to
build up. When the disordered layer reaches a critical size,
another large event is unleashed; this is the underlying reason
for the quasiperiodic form of the time series shown in Figure
3(a).

The {\it bulk} packing fraction $\phi$ of the disordered sandpile
 increases
after a large event, i.e. effective consolidation occurs during
avalanching.  Internal consolidation 
and surface smoothing are, therefore, closely related.
Also, a comparison of the surface width for  pre- and post- large event sandpiles
 shows that the surface width goes down
considerably during an event, once again suggesting that a rough
precursor pile is smoothed by the propagation of a large
avalanche.

We turn finally
to the measurement of the
critical exponents defined above.

Our results are \cite{curr}:
\begin{itemize}
\item For disordered sandpiles ($L=2048$)
we find $\beta=0.42 \pm 0.05$; for ordered sandpiles ($L=2048$)
$\beta=0.17 \pm 0.05$.
\item For disordered sandpiles above a crossover
size of $L_c=90$
we find $\alpha=0.723 \pm 0.04$; while for ordered piles
we find $\alpha = 0.356 \pm 0.05$.
\item Based on the above values we find the dynamical
exponent $z$, has values of  $1.72 \pm 0.29$ and $2.09 \pm 0.84$
for the disordered and ordered sandpiles.
\end{itemize}

Since the effect of large avalanches is to transform
a disordered pile into a largely ordered one (Fig. 3),
we notice that the above exponents confirm the smoothing of the surface.
It is important to realise that it is the mechanism
of column reorganisation, our representation of the slow
dynamics of the system, that causes  the initial
accumulation of grains resulting in the roughness of the precursor surface,
and thus the eventual smoothing of the surface. We emphasise that the
addition of 
such slow dynamics, independent of model details,
   is expected to lead have similar consequences. For example, the crucial
role of the cooperative mechanism has also been confirmed
in recent analytical investigations of the asymptotic smoothing
of continuum sandpile surfaces \cite{amcoupled2};
it has also been seen to influence the geometrical features of
two-dimensional model avalanches \cite{curr}
(cf. recent experiments on sloping beds of spheres\cite{daerr}).
\section{Acknowledgments}
\label{acknowledgements}
 GCB acknowledges support from the BBSRC,
 UK ($218$/FO$6522$).


\begin{thebibliography}{999}
\bibitem{caparam} Disordered sandpiles
described here have
 parameters \cite{amca} $z_0=2, z_1=20$ and $a =
0.7$, unless otherwise stated.
\end{thebibliography}
\end{document}